\documentclass[reprint,superscriptaddress,preprintnumbers,nofootinbib,amsmath,amssymb,aps]{revtex4-2}

\usepackage{graphicx}% Include figure files
\usepackage{dcolumn}% Align table columns on decimal point
\usepackage{bm}% bold math
\usepackage{amssymb}    %Allows use math symbols
\usepackage{color}         %Allows {\color{red} Hello World} or \colorbox{blue}{Hello World}
\usepackage{graphicx}     %Allows\includegraphics{figure.pdf}
\usepackage{mathrsfs} %Allows \mathscr{HELLO}
\usepackage{hyperref} %Automatically links \label and \ref commands; Always load last
\hypersetup{colorlinks=true,linkcolor=blue,citecolor=blue}
\usepackage{ upgreek } % for uptau
\usepackage{color,xcolor,fancybox,epsf,rotating,colordvi}

\begin{document}

\preprint{The 32nd International Symposium on Lepton Photon Interactions at High Energies, \\ Madison, Wisconsin, USA, August 25-29, 2025}

\title{Higgs Boson CP Properties and Effective Field Theory Measurements from the ATLAS Experiment at the LHC}% Force line breaks with \\
\author{Haijun Yang (on behalf of the ATLAS Collaboration)}
%\email[]{haijun.yang@sjtu.edu.cn, haijun.yang@cern.ch}
\email[Corresponding author:]{haijun.yang@sjtu.edu.cn} 
\affiliation{School of Physics and Astronomy, Shanghai Jiao Tong University, \\  800 Dongchuan Road, Shanghai 200240,  China}
\affiliation{Tsung-Dao Lee Institute, Shanghai Jiao Tong University, \\ 1 Lisuo Road, Shanghai 201210, China}
\affiliation{State Key Laboratory of Dark Matter Physics, \\Key Laboratory for Particle Astrophysics and Cosmology (MOE), 
\\Shanghai Key Laboratory for Particle Physics and Cosmology (SKLPPC)}
% \collaboration{CLEO Collaboration}%\noaffiliation

\date{\today}% It is always \today, today,

\begin{abstract}

This proceedings presents a concise overview of the Higgs boson's charge-conjugation and parity (CP) properties and constraints on Effective Field Theory (EFT) operators, derived from the ATLAS experiment at the Large Hadron Collider (LHC). Using proton$\textendash$proton collision data with integrated luminosities of up to 140 fb$^{-1}$ at $\sqrt{s} = 13$ TeV, the ATLAS collaboration systematically probe the CP nature of the Higgs boson's couplings to fermions ($\tau$ leptons, bottom quarks, and top quarks) and bosons ($W$, $Z$, and $\gamma$) across diverse decay final states.
The EFT framework is used to parameterize Beyond the Standard Model (BSM) effects via dimension-6 operators, enabling model-independent constraints on CP violation and new physics scales. 
 The main focus is comparison of measurement characteristics and sensitivity across different final states: (1) $H \to \tau\tau$ (semileptonic/hadronic decays) for light fermion couplings; (2) $H \to \gamma\gamma$ and $H \to bb$ in $t\bar{t}H/tH$ processes for heavy fermion couplings; (3) $H \to WW^* \to l\nu l\nu$, $H \to ZZ^* \to 4l$, and vector boson fusion (VBF) $H \to \tau\tau/\gamma\gamma$ for boson couplings; and (4) double Higgs ($HH$) production for self-couplings. All measurements are consistent with the Standard Model (SM) prediction of a CP-even Higgs boson ($J^{CP} = 0^{++}$), with no evidence of CP violation. The most stringent constraint on the CP-odd EFT parameter $c_{H\tilde{W}}$ is obtained from VBF $H \to \tau\tau$ ($c_{H\tilde{W}} \in [-0.23, 0.70]$ at 95\% CL), highlighting the unique sensitivity of this channel. Complementary constraints from other final states reinforce the robustness of SM consistency and provide a foundation for future searches.
 \end{abstract}

\maketitle
\newpage

%\tableofcontents
% \newpage

\section{\label{sec:Introduction}Introduction}

The discovery of the Higgs boson at the LHC in 2012 by the ATLAS and CMS collaborations \cite{Aad:2012tfa,Chatrchyan:2012ufa} confirmed the Brout-Englert-Higgs mechanism for electroweak symmetry breaking, completing the SM of particle physics. In the SM, the Higgs boson is a pure CP-even scalar ($J^{CP} = 0^{++}$), with couplings to fermions and bosons strictly adhering to CP conservation \cite{Gunion:2000gf}. However, the SM fails to address fundamental cosmological puzzles—most notably the matter-antimatter asymmetry of the universe \cite{Cohen:1993nk}—which implies the existence of BSM physics that may introduce CP violation. Anomalous Higgs couplings, including CP-mixed components, are prime candidates for mediating such BSM effects, making the study of Higgs CP properties a critical frontier in particle physics.

Effective Field Theories (EFTs) provide powerful, model-independent frameworks to parameterize BSM effects at LHC-accessible energies. By extending the SM Lagrangian with higher-dimension operators (dominantly dimension-6 for LHC kinematics), EFTs allow us to constrain deviations from SM couplings without assuming a specific BSM model. The Wilson coefficients of these operators quantify BSM effect strength, and their constraints translate to bounds on the scale of new physics (typically $\Lambda = 1$ TeV).
This paper summarizes the latest ATLAS results on Higgs CP properties and EFT measurements from different Higgs couplings and decay final states. 
%Section \ref{sec:theory} reviews the theoretical framework for Higgs CP decomposition and EFT parameterization. Section \ref{sec:experiment} describes the ATLAS detector, experimental data, Monte Carlo (MC) simulations, and analysis techniques (including CP-sensitive observables). Sections \ref{sec:fermion}–\ref{sec:double_higgs} present results for Higgs-fermion couplings, Higgs-boson couplings, and double Higgs production, respectively, with detailed discussions of final-state-specific sensitivity. Section \ref{sec:discussion} synthesizes these results and compares sensitivities across channels. Section \ref{sec:summary} concludes with key findings and future prospects.

%%%%%%%%%%%%%%%%%%%%%%%%
\section{Theoretical Framework}
\label{sec:theory}

%\subsection{2.1 Higgs CP Decomposition}
CP symmetry implies that Higgs couplings to SM particles can be decomposed into orthogonal CP-even and CP-odd components. For a Higgs boson of mass $m_H = 125$ GeV, the CP nature is probed by measuring observables sensitive to interference between these components %\cite{Isidori:2001bm}.

%\subsubsection{2.1.1 Fermion Couplings ($H \to ff$)}
For fermion couplings (e.g., $H \to b\bar{b}$, $t\bar{t}$, $\tau\tau$) the interaction Lagrangian is parameterized as~\cite{EPJC:83:563}:
\begin{equation*}
\mathcal{L}_{Hff} = -\frac{m_f}{v} \kappa_f \left( \cos\phi_f \overline{f}f + \sin\phi_f \overline{f}i\gamma_5 f \right) H
\end{equation*}
where $v = 246$ GeV is the Higgs vacuum expectation value, $\kappa_f$ is the coupling strength relative to the SM, and $\phi_f$ is the CP-mixing angle. A pure CP-even coupling corresponds to $\phi_f = 0^\circ$ (only the scalar term $\overline{f}f$ contributes), while $\phi_f = 90^\circ$ indicates a pure CP-odd coupling (only the pseudoscalar term $\overline{f}i\gamma_5 f$ contributes).
%\cite{EPJC:83:563}. 
Deviations from $\phi_f = 0^\circ$ signal CP violation.

%\subsubsection{2.1.2 Boson Couplings ($H \to VV/\gamma\gamma$)}
For boson couplings (e.g., $H \to WW^*/ZZ^*$, $H \to \gamma\gamma$), CP sensitivity arises from interference between the SM CP-even amplitude and BSM CP-odd contributions. In the SM, $H \to VV$ (where $V = W/Z$) proceeds via $s$-channel vector boson fusion or $t$-channel fermion loops, both of which are CP-even. CP-odd contributions modify the $HVV$ vertex via BSM operators, leading to azimuthal or angular distributions distinct from the SM \cite{JHEP:05:105}.
For $H \to \gamma\gamma$, the SM amplitude is dominated by top-quark and $W$-boson loops (CP-even). CP-odd contributions (e.g., from BSM fermion loops or EFT operators) introduce phase shifts in the photon decay plane, which are measurable via angular observables \cite{PRL:131:061802}.

%\subsection{2.2 Effective Field Theory (EFT) Parameterization}
The SM Effective Field Theory (SMEFT) extends the SM Lagrangian with higher-dimension operators, suppressed by powers of the BSM scale $\Lambda$:
\begin{equation*}
\mathcal{L}_{\text{SMEFT}} = \mathcal{L}_{\text{SM}} + \sum_{i}^{N_{d6}} \frac{c_i}{\Lambda^2} O_i^{(6)} + \sum_{j}^{N_{d8}} \frac{b_j}{\Lambda^4} O_j^{(8)} + \dots,
\end{equation*}
where $c_i$ ($b_j$) are Wilson coefficients for dimension-6 (d=8) operators. For the processes discussed in this proceedings, dimension-6 operators dominate, as d=8 terms are suppressed by $1/\Lambda^4$.
%\subsubsection{2.2.1 Key EFT Bases}
Two bases are critical for Higgs CP and coupling measurements:
\begin{itemize}

\item \textit{Warsaw Basis}: Includes 16 dimension-6 operators relevant to Higgs physics, with CP-odd contributions mainly from $O_{H\tilde{W}}$ ($H^\dagger H \tilde{W}_{\mu\nu}^I W^{I\mu\nu}$), $O_{H\tilde{B}}$ ($H^\dagger H \tilde{B}_{\mu\nu} B^{\mu\nu}$), and $O_{H\tilde{W}B}$ ($H^\dagger \tau^I H \tilde{W}_{\mu\nu}^I B^{\mu\nu}$). These operators modify the $HVV$ vertex and introduce CP violation, with Wilson coefficients $c_{H\tilde{W}}$, $c_{H\tilde{B}}$, and $c_{H\tilde{W}B}$.

\item \textit{HISZ Basis}: Rotated to simplify Higgs-gauge boson couplings, with parameters $\tilde{d}$ and $\tilde{d}_B$ relating to $H \to \gamma\gamma$ and $H \to \gamma Z$ couplings. The $H \to \gamma\gamma$ coupling in this basis is:
  \begin{equation*}
  \mathcal{L}_{H\gamma\gamma} = \frac{g}{2m_W} \left( \tilde{d} \sin^2\theta_W + \tilde{d}_B \cos^2\theta_W \right) H \tilde{F}_{\mu\nu} F^{\mu\nu},
  \end{equation*}
  where $\theta_W$ is the weak mixing angle and $\tilde{F}_{\mu\nu}$ is the dual field strength tensor. The parameter $\tilde{d} = 0$ in the SM, so non-zero $\tilde{d}$ indicates BSM/CP-odd effects.
\end{itemize}

%\subsubsection{2.2.2 Sensitivity Enhancement via PCA}
%Wilson coefficients in the Warsaw basis are often correlated (e.g., multiple operators affect $H \to WW^*$ decays), leading to reduced fit convergence. To address this, Principal Component Analysis (PCA) is applied to rotate the basis into orthogonal linear combinations of Wilson coefficients, maximizing the analysis sensitivity to BSM effects \cite{JHEP:05:105}. The covariance matrix of the rotated coefficients is derived from the STXS (Simplified Template Cross Section) signal strength covariance matrix:
%\begin{equation}
%C_{\text{SMEFT}}^{-1} = P^\top C_{\text{STXS}}^{-1} P
%\end{equation}
%where $P_{ij} = A_j^{\sigma_i} + A_j^{\Gamma_{H\to WW^*}} - A_j^{\Gamma_H}$ (with $A_j^{\sigma_i}$ and $A_j^\Gamma$ denoting EFT corrections to production cross sections and decay widths).

%%%%%%%%%%%%%%%%%%%%%%%%
\section{ATLAS Experiment and Analysis Techniques}
\label{sec:experiment}

The ATLAS detector~\cite{ATLAS:2008instr} at the LHC covers nearly the entire solid angle around the collision point. It consists of an inner tracking detector surrounded by a thin superconducting solenoid, electromagnetic and hadronic calorimeters, and a muon spectrometer incorporating three large superconducting air-core toroidal magnets.
 
%\subsection{3.1 Data and Monte Carlo Simulations}
These analyses use LHC proton$\textendash$proton collision data collected by the ATLAS detector from Run 2 (2015$\textendash$2018) at $\sqrt{s} = 13$ TeV with integrated luminosity of $139$–$140$~fb$^{-1}$. 
For $H \to ZZ^* \to 4l$ analysis, data from Run 3 (2022$\textendash$2023) at $\sqrt{s} = 13.6$ TeV with integrated luminosity of $56$ fb$^{-1}$ are also used to increase statistical precision of the measurement~\cite{ATLAS-CONF-2025-002}.
The MC generators and orders of the calculations in the QCD and electroweak couplings for the signal and major background processes are listed in TABLE~\ref{tab:mc}.

\begin{table}[ht]
\centering
\caption{MC generators and theoretical orders for key signal and background processes~\cite{PLB:2020:805}.}
\label{tab:mc}
\begin{tabular}{lcc}
\toprule
Process &  Generator &  Theoretical Order \\ \hline
%\textit{Signal} & &  \\
ggF $H$ & POWHEG Box v2 & N$^3$LO QCD + NLO EW \\
VBF $H$ & POWHEG Box v2 & NNLO QCD + NLO EW \\
$VH$ & POWHEG Box v2 & NNLO QCD + NLO EW \\
$t\bar{t}H$ & MG5\_aMC@NLO & NLO QCD + NLO EW \\ \hline
%\textit{Background} & & \\
$t\bar{t}$ & POWHEG Box v2 & NNLO + NNLL \\
W/Z+jets & SHERPA 2.2.1 & NNLO \\
VV ($WW/ZZ$) & SHERPA 2.2.1 & NLO \\
EW W/Z+jj & SHERPA 2.2.1 & LO \\ \hline
\end{tabular}
\end{table}

%\subsection{3.2 CP-Sensitive Observables}

The choice of CP-sensitive observable is critical to a channel's sensitivity, with final state characteristics dictating optimal selections of observables:
\begin{itemize}
    \item $H \to \tau\tau$: The signed acoplanarity angle $\phi^*_{CP}$ measures the relative orientation of the two $\tau$ decay planes. For CP-even couplings, $\phi^*_{CP}$ is symmetric around 0°; for CP-odd couplings, it exhibits a forward-backward asymmetry~\cite{EPJC:83:563}.
  
    \item VBF $H \to \tau\tau/\gamma\gamma$: The jet azimuthal angle difference $\Delta\phi_{jj}$ (between the two VBF jets) and optimal observable (OO) maximize CP sensitivity. The OO is defined as:
  \begin{equation*}
  OO = \frac{2\mathfrak{R}\left(\mathcal{M}_{\text{SM}}^* \mathcal{M}_{\text{CP-odd}}\right)}{|\mathcal{M}_{\text{SM}}|^2}
  \end{equation*}
  where $\mathcal{M}_{\text{SM}}$ and $\mathcal{M}_{\text{CP-odd}}$ are the SM and CP-odd amplitudes. The OO provides a 30\% stronger constraint than $\Delta\phi_{jj}$ by leveraging interference effects \cite{arXiv:2506.19395}.

  \item $H \to ZZ^* \to 4l$: The production$\textendash$level and decay$\textendash$level observable OO are sensitive to CP-odd contributions~\cite{JHEP:05:105}.
  
  \item $t\bar{t}H \to bb$: Observables $b_2$ and $b_4$ quantify the azimuthal separation and beamline alignment of the top quarks. The observable $b_2$ is enhanced for narrow azimuthal separation in CP-odd scenarios, while $b_4$ is enhanced for top quarks aligned with beamline \cite{PLB:849:138469}.
\end{itemize}

%\subsection{3.3 Likelihood Fitting and EFT Interpretation}

All analyses use binned maximum likelihood fits to extract CP-mixing angles ($\phi_f$) or Wilson coefficients ($c_i$). Statistical and systematic uncertainties are included
%- \textit{Statistical}: From limited data/MC statistics.
%- \textit{Experimental}: Detector resolution, jet/lepton identification, and missing transverse momentum ($ {E}_T$) reconstruction.
%- \textit{Theoretical}: Signal cross section uncertainties, PDF variations, and parton shower modeling.
For EFT interpretations, we assume $\Lambda = 1$ TeV and include linear (and, where relevant, quadratic) terms in Wilson coefficients to account for higher-order EFT effects ~\cite{arXiv:2504.07686}. Fits are performed in 1D (single coefficient) and 2D (correlated coefficients) to quantify constraints.

%%%%%%%%%%%%%%%%%%%%%%%%
\section{Higgs CP Properties: Couplings to Fermions}
\label{sec:fermion}

\subsection{$H \to \tau\tau$: Light Fermion CP}

The $H \to \tau\tau$ channel is unique in its sensitivity to light fermion couplings, with two decay modes analyzed. 
%Semileptonic ($\tau_{\text{lep}} \to e/\mu + \nu$ and $\tau_{\text{had}} \to \pi/\rho + \nu$) and hadronic ($\tau_{\text{had}} \to \tau_{\text{had}}$)~\cite{EPJC:83:563}.
%\subsubsection{4.1.1 Analysis Details}
Using 139 fb$^{-1}$ of Run 2 data, events are selected with two $\tau$ candidates (leptonic/hadronic) and $E_{\textup{T}} > $~20 GeV. Backgrounds include $Z/\gamma^* \to \tau\tau$ (dominant), $t\bar{t}$, and QCD multijets, which are modeled via MC and data-driven methods. 
The CP-sensitive observable $\phi^*_{CP}$ is reconstructed as the angle between the two $\tau$ decay planes, weighted by their momentum. The $\phi^*_{CP}$ directly probes the pseudoscalar component of the $H\tau\tau$ coupling, with minimal theoretical uncertainties (due to well-measured $\tau$ decay properties).
A 1D likelihood fit to $\phi^*_{CP}$ yields the CP-mixing angle of 
$\phi_\tau = 9^\circ \pm 16^\circ$ (68\% CL). The pure CP-odd hypothesis ($\phi_\tau = 90^\circ$) is excluded at $3.4\sigma$, with the coupling strength $\kappa_\tau = 1.01 \pm 0.07$ (68\% CL) consistent with the SM \cite{EPJC:83:563}.

\subsection{$ttH, H \to \gamma\gamma, b\bar{b}$: Heavy Fermion CP}

The Higgs$\textendash$top quark coupling is probed via $t\bar{t}H$ and $tH$, with $H$ decaying to $\gamma\gamma$ or $bb$, two final states with complementary sensitivities.

%\subsubsection{4.2.1 $t\bar{t}H \to \gamma\gamma$}
%- \textit{Final-State Characteristics}: 
$H \to \gamma\gamma$ decays have excellent invariant mass resolution, making them ideal for precise coupling measurements. However, $t\bar{t}H$ production has a small cross section ($\sigma_{t\bar{t}H} \approx 0.5$ pb at 13 TeV), limiting the statistical precision of ${t\bar{t}H}$ measurements \cite{PRL:125:061802}.
%- \textit{Analysis Details}: 
Using 139 fb$^{-1}$ of Run 2 data, events are selected with two photons ($m_{\gamma\gamma} \in [120, 130]$ GeV), two top quark candidates, and $E_{\textup{T}} > $ 40~GeV. Major background includes $t\bar{t} + \gamma\gamma$.
%- \textit{Sensitivity and Results}: 
$t\bar{t}H$ is observed at $5.2\sigma$ assuming CP-even coupling. A fit to the CP-mixing angle $\alpha$ excludes the pure CP-odd hypothesis ($\alpha = 90^\circ$) at $3.9\sigma$, with $|\alpha| > 43^\circ$ excluded at 95\% CL \cite{PRL:125:061802}. The high purity of $\gamma\gamma$ decays makes this the most precise channel for $ttH$ CP measurement.

%\subsubsection{4.2.2 $t\bar{t}H/tH \to bb$}
%- \textit{Final-State Characteristics}: 
$H \to bb$ measurements have higher statistical precision than $H \to \gamma\gamma$ due to the large branching ratio $\mathcal{B}(H \to bb) \approx 58\%$, but suffer from huge $t\bar{t} + b$-jet background. This requires advanced background subtraction and CP-sensitive observables tailored to top quark kinematics \cite{PLB:849:138469}.
%- \textit{Analysis Details}: 
Using 139 fb$^{-1}$ of data, events are selected with two $b$-jets ($m_{bb} \in [110, 140]$ GeV), top quark candidates (leptonic/hadronic), and $\geq 4$ jets. CP sensitivity is derived from $b_2$ (azimuthal separation of top quarks) and $b_4$ (beamline alignment).
%- \textit{Sensitivity and Results}: 
The best-fit CP-mixing angle and coupling strength are:
  \begin{equation*}
  \alpha = 11_{-73^{\circ}}^{+52^{\circ}}, 
  \quad \kappa_t' = 0.84_{-0.46}^{+0.30} \quad (68\% \text{ CL}).
  \end{equation*}
  The pure CP-odd hypothesis is disfavored at only $1.2\sigma$, reflecting larger background-induced uncertainties~\cite{PLB:849:138469}.

%%%%%%%%%%%%%%%%%%%%%%%%
  
\section{Higgs CP Properties and EFT Constraints: Couplings to Bosons}
\label{sec:boson}
HVV couplings ($H\to WW^*/ZZ^*/\gamma\gamma$) are probed across diverse final states, with each channel offering unique sensitivity to EFT operators and CP violation.

\subsection{$H \to WW^* \to l\nu l\nu$}
The $H \to WW^* \to l\nu l\nu$ channel (where $l = e/\mu$) is the most statistically powerful Higgs decay mode ($\mathcal{B}(H \to WW^*) \approx 21\%$) but suffers from missing momentum due to the presence of neutrinos.
%\subsubsection{5.1.1 Analysis Details}
Using 140 fb$^{-1}$ of Run 2 data, events are selected with two leptons (opposite charge), $E_{\textup{T}} > $ 30~GeV, and no high-$p_{\textup{T}}$ jets (ggF) or two VBF jets. Backgrounds include $WW$ (dominant), $t\bar{t}$, and $ZZ$, which are all modeled using MC simulation. CP sensitivity is derived from $\Delta\phi_{jj}$ (VBF) and STXS categorization (binned by $p_{\textup{T}}^H$ and jet multiplicity) \cite{arXiv:2504.07686}.

%\subsubsection{5.1.2 EFT Constraints and Sensitivity}
EFT fits include 16 Warsaw-basis dimension-6 operators, reduced to five orthogonal operators. The result is compatible with the SM with a $p$-value of 91\%.
For the CP-odd Wilson coefficient, the fitted constraint is:
  \begin{equation*}
  c_{H\tilde{W}} \in [-1.0, 0.6] \quad (95\% \text{ CL}).
  \end{equation*}
The measured fiducial cross section is $\sigma_{\text{ggF}} \times \mathcal{B}(H \to WW^*) = 12.4_{-1.2}^{+1.3}$ pb which is consistent with SM prediction of 12.1 pb ~\cite{arXiv:2504.07686}.

%\textit{Sensitivity Characteristics}: 
%High statistics but moderate purity (due to $ {E}_T$ uncertainties). STXS categorization mitigates this by separating signal-rich regions, making this channel a workhorse for EFT fits.

\subsection{VBF $H \to \tau\tau$}
VBF production ($\sigma_{\text{VBF}} \approx 3.8$ pb) provides a clean Higgs sample with two forward jets, and the $H \to \tau\tau$ decay adds CP sensitivity via $\phi^*_{CP}$ and the OO.
%\subsubsection{5.2.1 Analysis Details}
Using 140 fb$^{-1}$ of Run 2 data, events are selected with two VBF jets ($m_{jj} > 500$~GeV, $\eta_j > 2.5$), two $\tau$ candidates, and $E_{\textup{T}} > $ 20~GeV. The OO is used to maximize CP sensitivity, with backgrounds modeled via MC and data-driven methods. The OO provides a 30\% stronger constraint than $\Delta\phi_{jj}$ by leveraging interference effects~\cite{arXiv:2506.19395}.
%\subsubsection{5.2.2 EFT Constraints and Sensitivity}
%- \textit{Key Result}: 
The most stringent constraint on $c_{H\tilde{W}}$ to date is obtained:
  \begin{equation*}
  c_{H\tilde{W}} \in [-0.23, 0.70] \quad (95\% \text{ CL})
  \end{equation*}
  No CP violation is observed ($p$-value = 99\%). VBF production has lower statistics than ggF, but the improved purity dominates, making this the most sensitive channel for $HVV$ CP measurement~\cite{arXiv:2506.19395}.
  
%- \textit{Sensitivity Characteristics}:
%  - \textit{Strengths}: 
%VBF jet tagging reduces background (e.g., $Z/\gamma^* \to \tau\tau$) by 50\% compared to inclusive $H \to \tau\tau$. The OO leverages interference between SM and CP-odd amplitudes, providing a 30% stronger constraint than $\Delta\phi_{jj}$.

\subsection{VBF $H \to \gamma\gamma$}

VBF $H \to \gamma\gamma$ combines the high resolution of $\gamma\gamma$ decays with the clean production of VBF, making it sensitive to $H\gamma\gamma$ CP properties.
%\subsubsection{5.3.1 Analysis Details}
Using 139 fb$^{-1}$ of Run 2 data, events are selected with two photons ($m_{\gamma\gamma} \in [118, 132]$~GeV), two VBF jets ($m_{jj} > 600$~GeV), and $ {E}_{\textup{T}} < $ 20~GeV. Backgrounds include $QCD~ \gamma\gamma$ (dominant) and $ggF~ H \to \gamma\gamma$, both of which are modeled via data-driven sidebands.
%\subsubsection{5.3.2 EFT Constraints and Sensitivity}
The CP-odd parameters $c_{H\tilde{W}}$ (Warsaw basis) and $\tilde{d}$ (HISZ basis) are constrained to:
  \begin{align*}
    c_{H\tilde{W}} \in [-0.53, 1.02] \quad (95\% \text{ CL}) \\
    \textup{and} ~ \tilde{d} \in [-0.032, 0.059] \quad (95\% \text{ CL}).
  \end{align*}

%- \textit{Sensitivity Characteristics}: 
Excellent photon resolution enables precise coupling measurements, but low VBF statistics ($\sigma_{\text{VBF}} \times \mathcal{B}(H \to \gamma\gamma) \approx 0.008$ pb) limits the overall sensitivity. This channel is unique in probing $H\gamma\gamma$ CP properties \cite{PRL:131:061802}.

\subsection{$H \to ZZ^* \to 4l$}

The $H \to ZZ^* \to 4l$ channel (where $l = e/\mu$) has the lowest background of any Higgs decay mode ($\mathcal{B}(H \to ZZ^*) \approx 2.6\%$) due to its distinctive four-lepton final state.
%\subsubsection{5.4.1 Analysis Details}
Using 139 fb$^{-1}$ Run 2 data, events are selected with four leptons (two pairs of opposite charge/same flavor), $m_{4l} \in [115, 130]$ GeV, and no high-$p_{\textup{T}}$ jets. Backgrounds are consisting of $ZZ^*$ and $t\bar{t}$ \cite{JHEP:05:105}.
%\subsubsection{5.4.2 EFT Constraints and Sensitivity}
The EFT fits to CP-odd operators yield constraints of three Wilson coefficients: 
  \begin{align*}
  c_{H\tilde{B}} \in [-0.61, 0.54] \quad (95\% \text{ CL}), \\
  c_{H\tilde{W}B} \in [-0.97, 0.98] \quad (95\% \text{ CL}), \\ 
  \textup{and} ~ c_{H\tilde{W}} \in [-0.81, 1.54] \quad (95\% \text{ CL}).
  \end{align*}
  All coefficients are consistent with the SM ($p$-value = 86\%).
With 56 fb$^{-1}$ of Run 3 data collected at $\sqrt{s}=$ 13.6~TeV, the measured fiducial cross section is:
  $\sigma_{\text{fid}} = 3.5 \pm 0.5 \, (\text{stat.}) ^{+0.3}_{-0.2} \, (\text{sys.}) \, \text{fb}$, which is consistent with the SM prediction of $3.7 \pm 0.1$ fb ~\cite{ATLAS-CONF-2025-002}.

%- \textit{Sensitivity Characteristics}: 
The low background enables precise EFT fits with minimal systematic uncertainties. However, the low branching fraction of $H \to ZZ^* \to 4l$ limits the statistical precision.

\subsection{$WH$, $H \to bb$}
The $WH$ production mode ($\sigma_{WH} \approx 1.3$ pb) with $H\to~bb$ decays probes $HVV$ couplings in a different production topology than ggF/VBF.
%\subsubsection{5.5.1 Analysis Details}
Using 140 fb$^{-1}$ of Run 2 data, events are selected with one lepton (from the $W$ boson decay), two $b$-jets ($m_{bb} \in [110, 140]$ GeV), and $ {E}_{\textup{T}} > $~30 GeV. A BDT separates signal from backgrounds ($t\bar{t}$, $W + b$-jets)~\cite{ATL-PHYS-PUB-2025-022}.
%\subsubsection{5.5.2 EFT Constraints and Sensitivity}
%- \textit{Key Result}: 
The CP-odd Wilson coefficient constraints:
  \begin{equation*}
  c_{H\tilde{W}} \in [-0.62, 0.85] \quad (95\% \text{ CL})
  \end{equation*}
  The measured signal strength is $\mu_{WH}^{bb} = 0.93_{-0.21}^{+0.22}$ (68\%~CL), consistent with the SM.
%- \textit{Sensitivity Characteristics}: 
WH production provides complementary sensitivity to $HVV$ couplings, as it is less affected by ggF-specific BSM effects. However, the $b$-jet background introduces systematic uncertainties, leading to weaker constraints than VBF $H \to \tau\tau$ ~\cite{ATL-PHYS-PUB-2025-022}.

\section{Double Higgs Production in EFT}
\label{sec:double_higgs}

Double Higgs ($HH$) production probes the Higgs self-coupling ($\lambda_{HHH}$), a critical parameter for electroweak symmetry breaking. $HH \to bbbb$, $bb\tau\tau$, and $bb\gamma\gamma$ final states are analyzed to constrain anomalous self-couplings using HEFT (Higgs EFT).
%\subsection{6.1 Analysis Details}
Using Run 2 (140 fb$^{-1}$) and Run 3 (168 fb$^{-1}$) data, events are selected for three final states. HEFT fits constrain two anomalous couplings:
  \begin{align*}
  c_{gghh} \in [-0.38, 0.49] \quad (95\% \text{ CL}) \\
 \textup{and} ~ c_{tthh} \in [-0.19, 0.70] \quad (95\% \text{ CL}).
  \end{align*}
Both these results are consistent with the SM ($c_{gghh} = c_{tthh} = 0$) ~\cite{PRL:133:101801, ATLAS-CONF-2025-005}
For the Run 2 and Run 3 combined results, 1D constraints are comparable to Run 2 alone, but 3D fits reveal significant correlations between $c_{gghh}$, $c_{tthh}$, and other HEFT operators. This highlights the need for multi-dimensional analyses to disentangle BSM effects. $HH$ production is uniquely sensitive to the Higgs self-coupling, but extremely low cross sections ($\sigma_{HH} \approx 32$ fb at 13 TeV) limit the statistical precision of the results. Future measurements performed with data collected at the HL-LHC will improve sensitivity.

%%%%%%%%%%%%%%%%%%%%%%%%%%%%%%%%%%%%%
\section{Comparison}
\label{sec:comparison}

Table \ref{tab:sensitivity_summary} summarizes the key characteristics and sensitivities of the Higgs production and decay channels analyzed.
\begin{table*}[htbp]
\centering
\caption{Summary of channel characteristics and sensitivity to Higgs CP and EFTs.}
\label{tab:sensitivity_summary}
\begin{tabular*}{\linewidth}{@{}lccccl@{}}
\toprule
Channel & Final State & Luminosity & Key Observable & $\sigma \times \mathcal{B}$ (fb) & Sensitivity Highlight \\ \hline
$H \to \tau\tau$ & $\tau_{\text{lep}}/\tau_{\text{had}}$ & 139 fb$^{-1}$ & $\phi^*_{CP}$ & 3488 & $H\tau\tau$ CP; balances purity/statistics \\
$t\bar{t}H \to \gamma\gamma$ & $\gamma\gamma + t\bar{t}$ & 139 fb$^{-1}$ & $m_{\gamma\gamma}$ & 1.15 & Most precise $ttH$ CP measurement \\
$t\bar{t}H \to bb$ & $bb + t\bar{t}$ & 139 fb$^{-1}$ & $b_2, b_4$ & 292 & High statistics; moderate background \\
$H \to WW^* \to l\nu l\nu$ & $l\nu l\nu$ & 140 fb$^{-1}$ & $\Delta\phi_{jj}$, STXS & 554 & Highest statistics; workhorse for EFT \\
VBF $H \to \tau\tau$ & $\tau\tau + VBF$ jets & 140 fb$^{-1}$ & OO, $\Delta\phi_{jj}$ & 238 & Most stringent $c_{H\tilde{W}}$ constraint \\
VBF $H \to \gamma\gamma$ & $\gamma\gamma + VBF$ jets & 139 fb$^{-1}$ & OO, $\Delta\phi_{jj}$ & 8.6 & Sensitive to $H\gamma\gamma$ CP \\
$H \to ZZ^* \to 4l$ & $4l$ & 139+56 fb$^{-1}$ & OO & 6.6 & Cleanest final state; low background \\
WH, $H \to bb$ & $l + bb$ & 140 fb$^{-1}$ & $m_{bb}$, BDT & 170 & Complementary $HVV$ sensitivity \\
$HH \to bbbb/bb\tau\tau/bb\gamma\gamma$ & $bb$ + $bb/\tau\tau/\gamma\gamma$ & 140+168 fb$^{-1}$ & $m_{bb}$, $m_{\gamma\gamma}$ & 13 & Unique to Higgs self-coupling \\ \hline 
\end{tabular*}
\end{table*}

\begin{figure}[!htbp]
    \centering
    \includegraphics[width=0.9\linewidth]{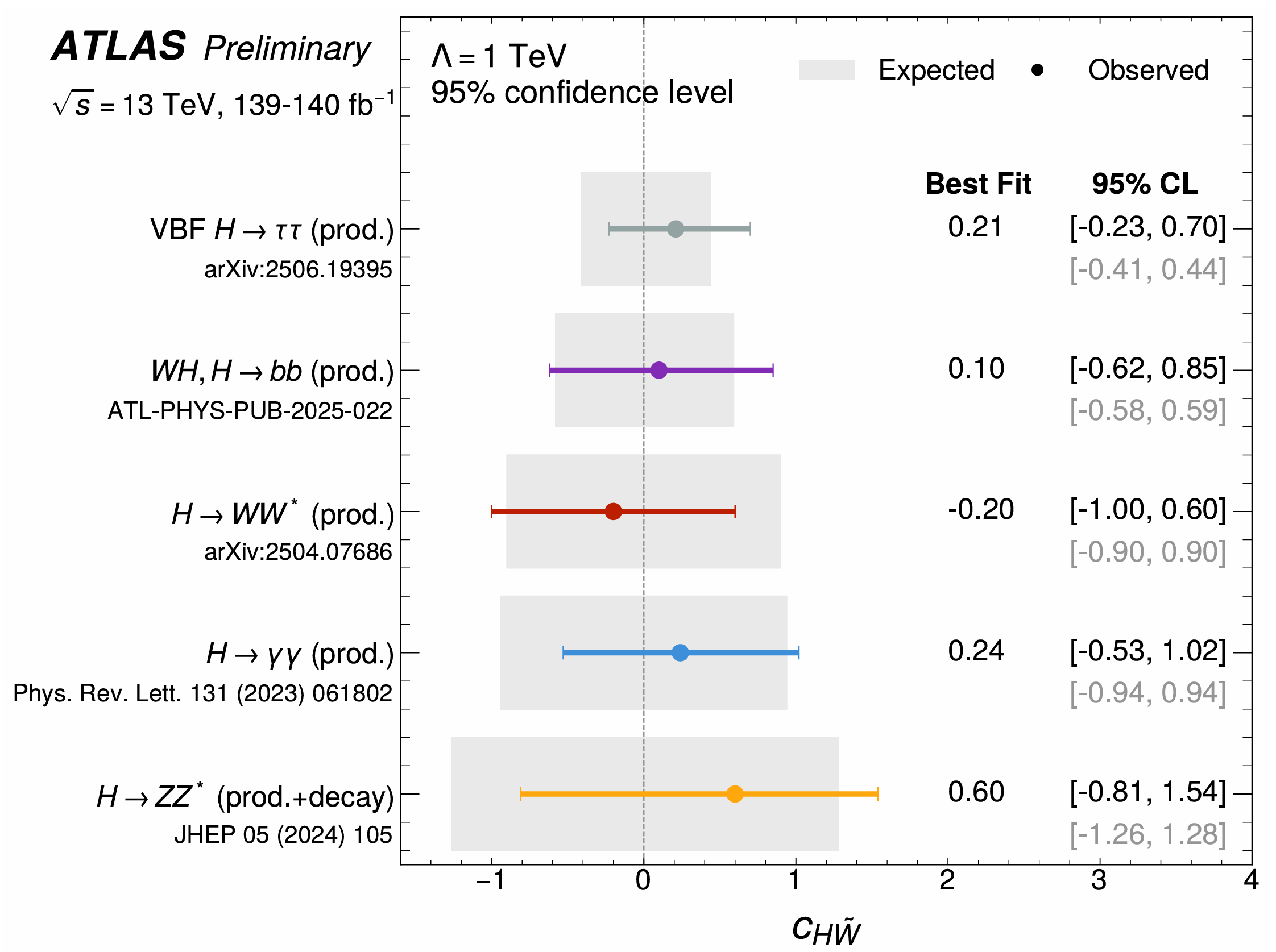}
    \caption{The expected and observed limits of the CP-odd Wilson coefficient $c_{H\tilde{W}}$ using ATLAS Run 2 data~\cite{ATL-PHYS-PUB-2025-031}.}
    \label{fig:ATLAS-EFT}
\end{figure}

Comparison of the results is made for the expected and observed measurements of the CP-odd Wilson coefficient $c_{H\tilde{W}}$ for an integrated luminosity of 140 fb$^{-1}$ at $\sqrt{s}$ = 13 TeV, as shown in FIG.~\ref{fig:ATLAS-EFT}. All results are obtained via a linear-only interpretation. All couplings scale as $1/\Lambda^2$ with the assumed value of $\Lambda$ = 1 TeV~\cite{ATL-PHYS-PUB-2025-031}.
 
%%%%%%%%%%%%%%%%%%%%%%%%
\section{Summary and Outlook}
\label{sec:summary}
This proceedings presents a comprehensive analysis of ATLAS measurements of Higgs CP properties and EFT constraints.
All measurements are consistent with the SM prediction of a CP-even Higgs boson ($J^{CP} = 0^{++}$). The pure CP-odd hypothesis is excluded at up to $3.9\sigma$ (for $t\bar{t}H \to \gamma\gamma$), with no evidence of CP violation across any channel.
The most stringent limit on the CP-odd Wilson coefficient $c_{H\tilde{W}}$ is $c_{H\tilde{W}} \in [-0.23, 0.70]$ (95\%~CL) from VBF $H \to \tau\tau$. All other EFT parameters are consistent with the SM. Each final state offers distinct strengths, VBF $H \to \tau\tau$ has the best CP sensitivity, $H \to ZZ^* \to 4l$ has the cleanest EFT fits, $H \to WW^* \to l\nu l\nu$ has the highest statistical precision, and $HH$ production provides unique self-coupling probe.

Future prospects will build on these results. The complete Run 3 dataset (projected 300 fb$^{-1}$) will reduce statistical uncertainties by ~40\%. Combining all channels into a single global fit will reduce Wilson coefficient correlations and enhance BSM sensitivity.
The HL-LHC (projected 3 ab$^{-1}$) will enable precision measurements of rare processes (e.g., $HH$ production).
These efforts will continue to test the SM's limits and search for new physics, advancing our understanding of electroweak symmetry breaking and CP violation.

%%%%%%%%%%%%%%%%%%%%%%%%
\begin{acknowledgments}
This work is supported by the National Natural Science Foundation of Chian (No. 1201101478) and the Ministry of Science and Technology of China (No. SQ2023YFA1600035). 
\end{acknowledgments}

%%%%%%%%%%%%%%%%%%%%%%%%%%%%%%%%%%%%%%%%
%\begin{thebibliography}{99}
%%\bibliographystyle{apsrev4-2}
\bibliography{ref} % Produces the bibliography via BibTeX.
%\include{ref.tex}
%\end{thebibliography}

\end{document}